\documentclass[useAMS,usenatbib]{mn2e}
\usepackage{graphicx}

\title[Timing and spectral study of XB 1254--690]{Timing and spectral study of XB 1254--690 using new {\it RXTE} PCA data}
\author[Mukherjee and Bhattacharyya]{Arunava Mukherjee$^{1}$\thanks{E-mail:
arunava@tifr.res.in} and Sudip Bhattacharyya$^{1}$\thanks{E-mail: sudip@tifr.res.in} \\
$^{1}$Department of Astronomy and Astrophysics, Tata Institute
of Fundamental Research, Mumbai 400005, India}

\begin{document}

\date{
}

\pagerange{\pageref{firstpage}--\pageref{lastpage}} \pubyear{2010}

\maketitle

\label{firstpage}
\begin{abstract}
We have analyzed the new {\it Rossi X-ray Timing Explorer} 
Proportional Counter Array data of the atoll neutron star (NS)
low-mass X-ray binary (LMXB) system XB 1254--690. 
The colour-colour diagram shows that the source was in 
the high-intensity banana state. We have found 
two low-frequency candidate peaks with single trial significances 
of $\approx 2.65 \times 10^{-8}$ and $\approx 7.39 \times 10^{-8}$ in the power 
spectra. After taking into account the number of trials,
the joint probability of appearance of these two peaks in the data set
only by chance is $\sim 4.5\times10^{-4}$, and hence a low-frequency QPO can be considered
to be detected with a significance of $\sim 4.5\times10^{-4}$, or, $\sim 3.5\sigma$
for the first time from this source. We have also done the first systematic 
X-ray spectral study of XB 1254--690, and 
found that, while one-component models are inadequate,
three-component models are not required by the data. We have
concluded that a combined broken-powerlaw and Comptonization model best describes the
source continuum spectrum among 19 two-component models.
The plasma temperature ($\sim 3$ keV) and the optical depth ($\sim 7$) of the
Comptonization component are consistent with the previously
reported values for other sources.
However, the use of a broken-powerlaw component to
describe NS LMXB spectra has recently been started, and we have
used this component for XB 1254--690 for the first time.
We have attempted to determine the relative energy budgets of
the accretion disc and the boundary layer using the best-fit spectral model, and concluded 
that a reliable estimation of these budgets requires correlations among
time variations of spectral properties in different wavelengths.
\end{abstract}


\section{Introduction}\label{Introduction}

The persistent neutron star low mass X--ray binary (LMXB) system XB 1254--690 
exhibits energy dependent intensity dips with the binary orbital 
period ($\sim 3.88$ hr), thermonuclear X-ray bursts,
and flares \citep{Griffithsetal1978, Courvoisieretal1986, 
Masonetal1980, Smaleetal2002}. This is a so-called 
``atoll" source \citep{Bhattacharyya2007}.
An atoll source typically traces a `C'-like curve in its colour-colour diagram 
or CCD \citep{vanderKlis2006}. The high-intensity lower part of this curve
is known as the banana state \citep{vanderKlis2006}. 
XB 1254--690 has always been found in the banana
state, and has never been observed in the spectrally hard low-intensity
island state \citep{Bhattacharyya2007}. Note that the position of a 
source on the CCD is usually correlated with the timing features observed
from the source \citep{vanderKlis2006}. Therefore, CCDs are useful to determine 
the source state, and hence to understand the related physics.

Fast brightness oscillations, i.e., burst oscillations, 
have been observed from thermonuclear bursts from several neutron star LMXBs
(\citet{StrohmayerBildsten2006}; \citet{Bhattacharyya2010} and references therein).
This timing feature provides one of the two ways to measure neutron star
spin frequencies for LMXB systems \citep{Chakrabartyetal2003, Bhattacharyya2010}.
\citet{Bhattacharyya2007} reported a suggestive evidence of 95 Hz
oscillations from a thermonuclear burst from XB 1254--690. This indicates that
the neutron star in this source spins with a frequency of 95 Hz, 
although further confirmation is required to establish this. 

Various types of quasi-periodic oscillations (QPOs) are observed from neutron star LMXBs
\citep{vanderKlis2006}. Their frequencies are roughly between millihertz (mHz) 
and kilohertz (kHz). Properties of one type of QPOs are sometimes correlated with
those of another, and typically these properties are correlated with the spectral
state inferred from CCD. Therefore, the QPOs can be very useful to understand
the accretion process around a neutron star. In addition, kHz QPOs can be 
useful to probe the strong gravity regime, and to measure the neutron star
parameters \citep{vanderKlis2006, Bhattacharyya2010}. However, any detection 
of a QPO from XB 1254--690 was never reported before this paper. 

The continuum spectrum of a neutron star LMXB is typically well described
with various models, giving different physical interpretations \citep{Linetal2007}.
For example, two classical spectral models, 
{\it Eastern} and {\it Western}, have been proposed for the high-intensity states 
\citep{Mitsudaetal1989, Whiteetal1988}. The {\it Eastern} model consists of a 
multicolour blackbody from the disc and a Comptonized component from the boundary
layer (the contact layer between the neutron star and the disc inner edge),
while the {\it Western} model has a Comptonized emission from the disc
and a single temperature blackbody from the boundary layer.
Although further detailed studies have been done with more sources
(e.g., \citet{ChurchBalucinskaChurch2001, ChristianSwank1997, MaccaroneCoppi2003,
MaitraBailyn2004, Gilfanovetal2003, Oliveetal2003, Wijnands2001, Barret2001}),
a general consensus on the appropriate X-ray spectral model has not been achieved.
It is therefore essential to do systematic spectral studies of
neutron star LMXBs using all the reasonable physical models. 
Even before going into the details, the first aim of 
such a study will be to identify which spectral component originates from the 
boundary layer, and which one comes from the disc, and hence to find out the
relative energy budgets of these two X-ray emitting components. 
These relative budgets not only will be useful to understand the accretion
process and components, but also will be helpful to constrain the neutron
star parameters and equation of state (EoS) models \citep{Bhattacharyyaetal2000,
Bhattacharyya2002, Bhattacharyya2010}. Recently,
\citet{Linetal2007} have done such a systematic study for two neutron star
LMXBs (Aql X--1 and 4U 1608--52). However, such a study has never been done
for XB 1254--690.

In this paper, we report the first evidence of a QPO from XB 1254--690. 
Moreover, we give the results and interpretations of the first systematic
X-ray spectral study of this source.
In \S~\ref{DataAnalysisandResults}, we mention the spectral and timing analysis
methods, and report the results. In \S~\ref{Discussion}, we discuss the interpretation
and importance of the observational results.

\section{Data Analysis and Results}\label{DataAnalysisandResults}

The neutron star LMXB system XB 1254--690 was 
observed with {\it Rossi X-ray Timing Explorer} ({\it RXTE}) between Jan 16, 
2008 (start time: 13:27:50; proposal no. P93062) and Mar 13, 2008 
(end time: 19:42:08) 
for a total observation time of 283.42 ks (total 60 observations; good-time after excluding 
bursts, dips, flares, datagaps, etc. is 257.39 ks). 
We have used the corresponding Proportional Counter Array (PCA) data set
for our analysis.
We have searched for quasi-periodic oscillations (QPOs) in this data set,
computed a CCD and a hardness-intensity diagram (HID),
and fitted the continuum spectra with various models extensively.
We have used Good Xenon data files for the timing analysis, and 
Standard2 (Std2) data files to perform spectral analysis and to compute the CCD
and the HID. 
  
\subsection{Colour-Colour and Hardness-Intensity Diagram} \label{Colour-ColourandHardness-IntensityDiagram}

\begin{figure*}
\centering
\includegraphics*[width=\textwidth]{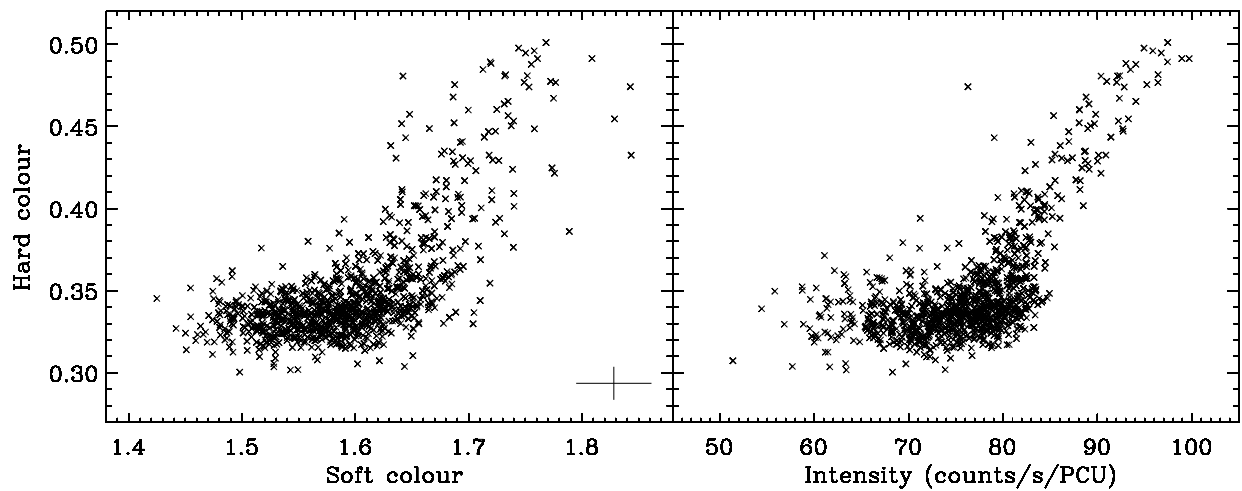}
\caption{Colour-colour diagram (CCD; left panel) and hardness-intensity diagram 
(HID; right panel) of the neutron star LMXB XB 1254--690. 
Here the soft-colour is defined as the ratio of the background 
subtracted counts in the energy range 5.7--7.5 keV to that in 4.4--5.7 keV, and 
the hard-colour is defined as the ratio of the background subtracted counts in 
11.4--20.7 keV to that in 7.5--11.4 keV. An HID is a plot of hard-colour vs intensity. 
The intensity is defined as the total background subtracted counts in the energy 
range 4.4--20.7 keV. These photon-counts are calculated from the Std2 data of 
the PCU2 unit of {\it RXTE} PCA. Typical error bars on a soft-colour and a hard-colour
are shown. This figure shows that the source was in the banana state during all the
observations (\S~\ref{Colour-ColourandHardness-IntensityDiagram}; see 
\citet{vanderKlis2006}).
\label{CCD1}}
\end{figure*}

We have shown a CCD and an HID of XB 1254--690 in Fig.~\ref{CCD1}. 
A CCD is a plot of a hard-colour vs a 
soft-colour. We have defined the soft-colour as the ratio of the 
background subtracted counts in the energy range 5.7--7.5 keV to that in 
4.4--5.7 keV. Similarly, the hard-colour has been defined as the ratio of the background 
subtracted counts in 11.4--20.7 keV to that in 7.5--11.4 keV. 
An HID is a plot of hard-colour vs intensity. The intensity has been defined as 
the total background subtracted counts in the energy range 4.4--20.7 keV. 
These photon-counts have been calculated from the Std2 data of the {\it RXTE} PCA instrument.
We have used only the PCU2 data, because this Proportional Counter Unit (PCU)
operated during all the observations.
Furthermore, we have used only the upper layers of PCU2 in order to get
a better signal-to-noise ratio. 

Fig.~\ref{CCD1} shows that the source was in the banana state during all the observations.
This is consistent with the earlier results (\S~\ref{Introduction}; 
see \citet{Bhattacharyya2007}).
We have detected two thermonuclear X-ray bursts, and several flares and dips in the
lower banana state. The bursts have not exhibited signatures of photospheric radius 
expansion. In addition, we have not detected burst oscillations from these bursts.

\subsection{Timing Analysis}\label{TimingAnalysis}

\begin{figure*}
\centering
\begin{tabular}{lr}
\hspace{-1.0cm}
\includegraphics*[width=.5\textwidth]{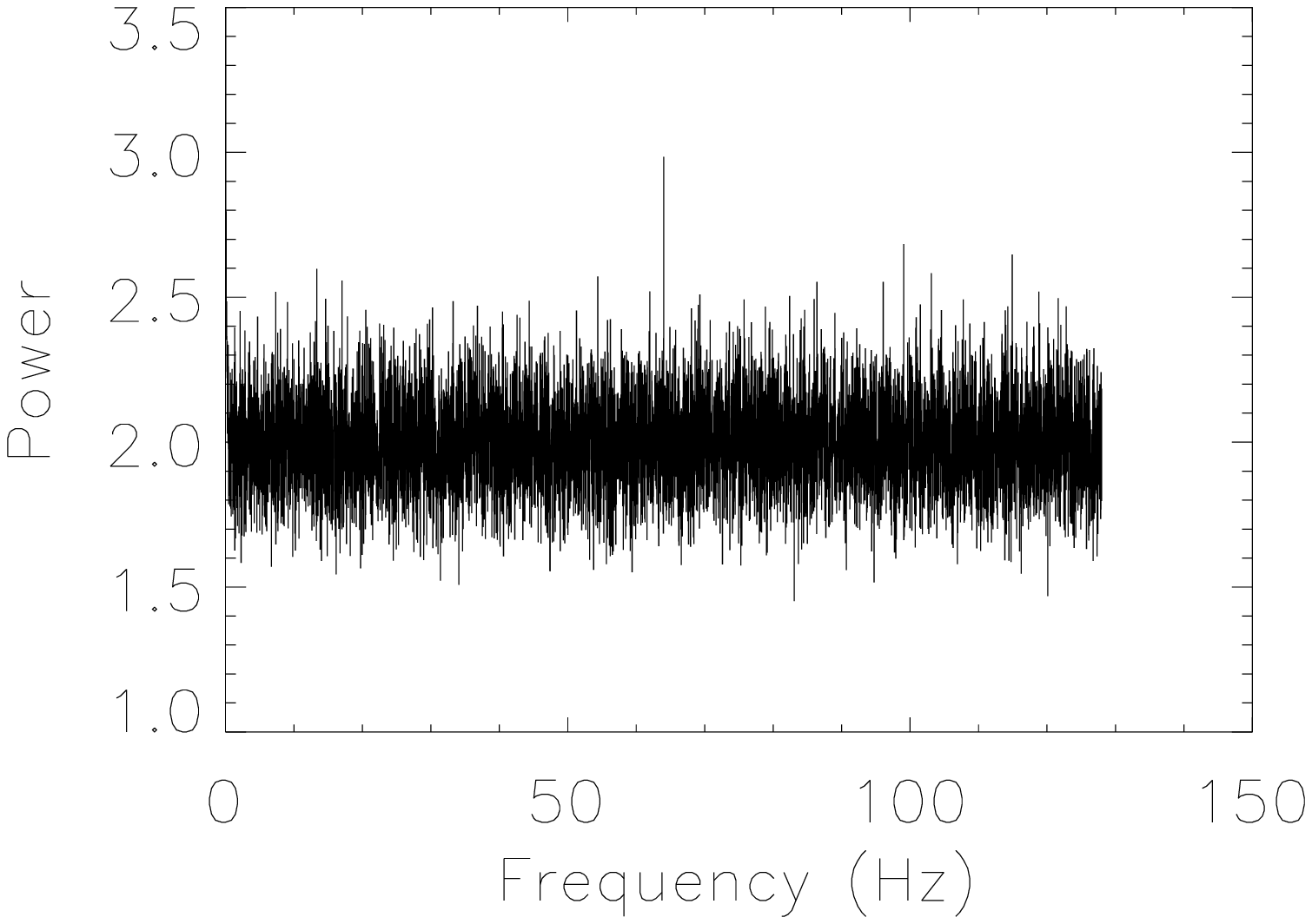} &
\includegraphics*[width=.5\textwidth]{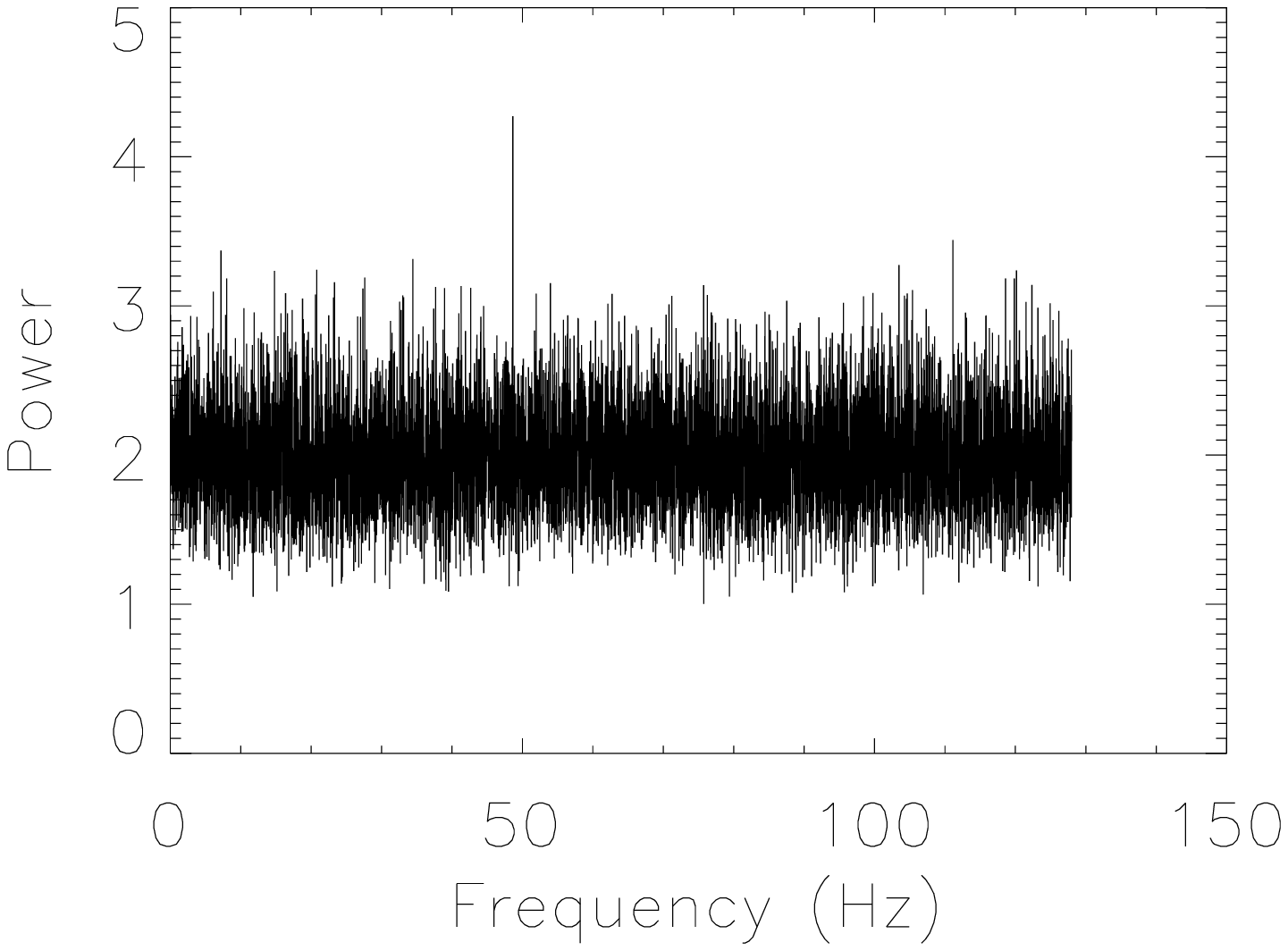}
\end{tabular}
\caption{Power spectra with frequency resolution of 0.016 Hz from {\it RXTE} PCA data of 
the neutron star LMXB XB 1254--690. 
{\it Left panel:} this spectrum corresponds to the observation done on Jan 21, 2008 (time:
10:44:08 to 14:49:08). A candidate peak at 64.01 Hz is clearly seen.
{\it Right panel:} this spectrum corresponds to the observation done on Jan 17-18, 2008
(time: 22:47:28 to 00:02:10). A candidate peak at 48.63 Hz is clearly seen.
The probability of appearence of these two peaks in our data set 
only by chance is $\sim 4.5\times10^{-4}$, and hence a low-frequency QPO can be considered
to be detected with a significance of $\sim 3.5\sigma$.
(\S~\ref{TimingAnalysis}).
\label{Powspec1}}
\end{figure*}

\begin{figure*}
\centering
\includegraphics*[width=\textwidth]{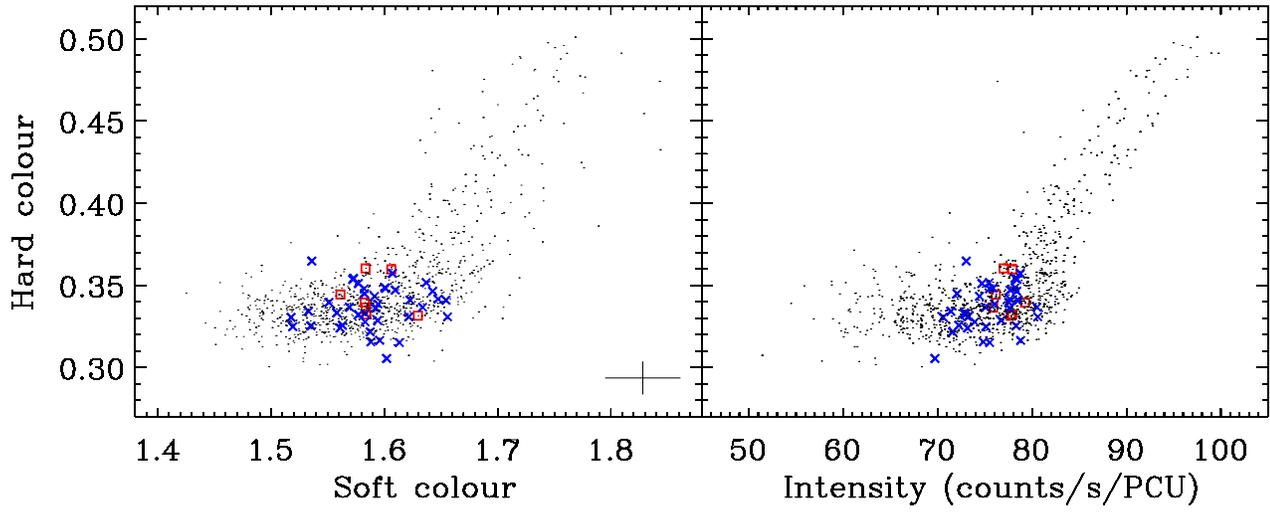}
\caption{Colour-colour diagram (CCD; left panel) and hardness-intensity diagram
(HID; right panel) of the neutron star LMXB XB 1254--690 (same as Fig.~\ref{CCD1}).
The data corresponding to the 64.01 Hz candidate QPO and 48.63 Hz candidate QPO 
are marked with blue {\it cross} signs and red {\it square} signs
respectively. This figure shows that both the candidate QPOs have appeared in the lower-banana
(LB) state (\S~\ref{TimingAnalysis}).
\label{CCD3}}
\end{figure*}

We have not found any significant kHz QPO in the entire {\it RXTE} PCA data set.
In order to search for low-frequency QPOs, we have calculated power spectra
(for the entire PCA energy range) 
with a Nyquist frequency of 128 Hz for 250 s intervals, 
using the Good Xenon data from PCU2. We have found two candidate peaks
around the frequency  $\sim 50$~Hz. The stronger one has appeared
at 64.01 Hz in the power spectrum corresponding to the observation done
on Jan 21, 2008 (time: 10:44:08 to 14:49:08). We have got a peak power of 
2.983, after averaging over forty 250 s intervals and merging four consecutive
frequency bins (see Fig.~\ref{Powspec1}, left panel). The latter has made the frequency 
resolution 0.016 Hz. The probability of obtaining a power
this high in a single trial from the expected $\chi^2$ noise
distribution (320 dof) is $\approx 2.65 \times 10^{-8}$. 
In order to find this peak, we have searched 60 power spectra, each with
8000 frequencies. Therefore, multiplying
with a number of trials of 480000, we have found a significance of $1.27\times10^{-2}$, 
or $\approx 2.5 \sigma$. The rms amplitude of this candidate QPO is 
$\approx 0.83$\%. The other candidate peak of same frequency-width and of 
$\approx 7.39 \times 10^{-8}$ single trial significance has been
found at 48.63 Hz from the observations on 
Jan 17-18, 2008 (time: between 22:47:28 to 00:02:10).
Its rms amplitude is $\approx 1.3$\% (Fig.~\ref{Powspec1}, right panel), and significance 
is $3.55\times10^{-2}$ after multiplying with the number of trials of 480000.
In Fig.~\ref{CCD3}, we have marked the data corresponding to the two candidate
QPOs in the CCD and the HID. This figure shows that both the candidate QPOs have 
appeared in the lower-banana (LB) state. 

With these two candidate peaks, next we ask the question what would be the
probability of having two such peaks simultaneously in our data set containing
60 power spectra only by chance. The 480000 powers mentioned in the previous paragraph 
are expected to be independent of each other, otherwise the significance of each peak would
increase by the decrease of the number of trials. Therefore, naively the probability
that the two candidate peaks would appear simultaneously in our data set only
by chance is $1.27\times10^{-2}\times3.55\times10^{-2} = 4.5\times10^{-4}$.
However, this number is expected to be an upper limit, i.e., this significance
is expected to be better for the following reason.
Suppose, there are $10^5$ data sets like ours. Then, on average, the first peak 
should appear in 1270 data sets, and the second peak should appear in 3550 data sets.
Now we ask the question: in how many data sets at least one power would exceed
the first candidate peak power (i.e., the larger power) 
and at least two powers would exceed the second candidate peak power
(i.e., the smaller power). This number will be less than 45, which is the
above mentioned naively expected value, because there would be a chance that 
in some of the data sets more than one power would exceed
the first candidate peak power and/or more than two powers would exceed the
second candidate peak power.
If this is true then the $4.5\times10^{-4}$ would be the conservative value 
of the significance of simultaneous appearence of the two candidate peaks
in our data set. We have verified this logical conclusion using simulated 
data with a pure Poisson process in the next paragraph. 

Our simulated time series and power spectra 
have been rigorously tested, and have passed the following tests among
others: (1) the counts in 1/4096 s bins of the time series follow the
Poisson statistics extremely well; (2) when the simulated Leahy power spectra are fitted 
with a constant, the best-fit value is consistent with 2 (an example value
is $1.9995\pm0.0014$ for a 1 ks time series with the observed count rate); 
(3) the cumulative power distribution of the simulated power spectra
matches well with $\chi^2$ distribution with the appropriate degrees of freedom.
We have generated 10 Ms of simulated data with the observed count rate, 
which has been divided into 10 groups,
each with 1000 time series of 1 ks. 
Note that, while counting within one group gives a random value, 
the sample of 10 groups gives the mean and the standard deviation.
We have assumed two values of power: $p_1$ and $p_2$ ($p_1 > p_2$),
and estimated the number ($N$) of 1 ks time series (within one group)
with at least one power above $p_1$ and at least two powers above $p_2$
in the three following ways.
(1) In the first way we have used the conservative significance mentioned 
in the previous paragraph. We have calculated the significance ($S1_{p1}$ and
$S1_{p2}$ respectively) of $p_1$ and $p_2$ in a 1 ks time series.
Then with $S1_{\rm joint} = S1_{p1}\times S1_{p2}$, the number $N$ would
be $N_{\rm conservative} = 1000 S1_{\rm joint}$.
(2) In the second way, we have taken into account the fact (mentioned in 
the previous paragraph) that $p_1$ may be exceeded
more than once, and/or $p_2$ may be exceeded more than twice in a 1 ks time
series. Therefore we have counted the number of 1 ks time series in which
$p_1$ is exceeded, and divided this number with 1000 to estimate the significance
$S2_{p1}$ of $p_1$. Similarly we have estimated the significance $S2_{p2}$ of $p_2$.
Then with $S2_{\rm joint} = S2_{p1}\times S2_{p2}$, the number $N$ would
be $N_{\rm realistic} = 1000 S2_{\rm joint}$.
(3) In the third way, we have simply counted the number $N = N_{\rm counted}$.
For various combinations of $p_1$ and $p_2$ values, we have found that $N_{\rm counted}$
is always consistent with $N_{\rm realistic}$, but always significantly less
than $N_{\rm conservative}$. This verifies the logical conclusion of the previous
paragraph, that $4.5\times10^{-4}$ is the conservative value
of the significance of simultaneous appearence of the two candidate peaks
in our data set. 

Note that the single trial significances (typically, say, $1.25\times10^{-5}$
and $5.0\times10^{-5}$) of our assumed 
$p_1$ and $p_2$ values are much worse than those of the candidate peaks.
We have assumed such powers, because, (1) much larger powers would not change our conclusions;
and (2) a similar simulation with the significances of the candidate peaks
would need $\sim 3\times10^6$ Ms simulated data, which would require
$\sim 10^8$ GB of computer disc space and an unrealistic amount of time.
Therefore, since the probability of the appearence of both the candidate peaks in our data set
only by chance is less than $4.5\times10^{-4}$, one of them can be a signal with
the significance of $\sim 4.5\times10^{-4}$. Since both of them are of low-frequency, 
a low-frequency QPO can be considered to be detected from XB 1254-690 
with a significance of $\sim 4.5\times10^{-4}$, or, $\sim 3.5 \sigma$.

\subsection{Spectral Analysis}\label{SpectralAnalysis}

The neutron star LMXB XB 1254--690 was observed with several X-ray observatories, e.g.,
{\it BeppoSAX}, {\it Chandra (HETGS)}, {\it XMM-Newton}, 
{\it RXTE} (PCA) \citep{Iariaetal2001, Iariaetal2007, DiazTrigoetal2006, Smaleetal2002}.
In order to fit the continuum spectra, these papers reported  
the usage of a few {\ttfamily XSPEC} 
models, such as {\ttfamily diskbb + comptt} (or {\ttfamily compst}), 
{\ttfamily powerlaw + bbody}, {\ttfamily cutoffpl + bbody}. 
But a systematic analysis of the 
XB 1254--690 continuum spectra using all the reasonable XSPEC models was 
previously not done. 
As mentioned in {\S \ref{Introduction}}, \citet{Linetal2007}
have recently performed a systematic analysis using various 
continuum spectral models for two neutron star LMXBs. 
In this spirit, we have done the first systematic continuum 
spectral analysis for XB 1254--690.

Let us first discuss the currently understood X-ray components of neutron star LMXBs.
Such a source plausibly has two primary X-ray emitting regions: 
a geometrically thin accretion disc, and a boundary layer.
Each of the disc and the boundary layer is expected to be optically
thick, and hence to emit like a blackbody. However, note that the disc blackbody
should be a multicolour blackbody, as the blackbody temperature should be 
a function of the radial distance from the centre of the neutron star
\citep{Bhattacharyyaetal2000}. In addition, the disc and/or the boundary layer
may be covered by one or more coronae (hot electron-gas), which may reprocess 
(Comptonize) the emitted radiation. Therefore, depending on the nature of the
cover (e.g., partial versus full, optical depth), the observed spectrum
of each of the disc and the boundary layer may be a blackbody, or 
Comptonized, or a combination of both. In this paper, we have studied these various
possibilities systematically, in order to identify the nature of the
X-ray emitting components.

\begin{figure*}
\centering
\includegraphics*[width=\textwidth]{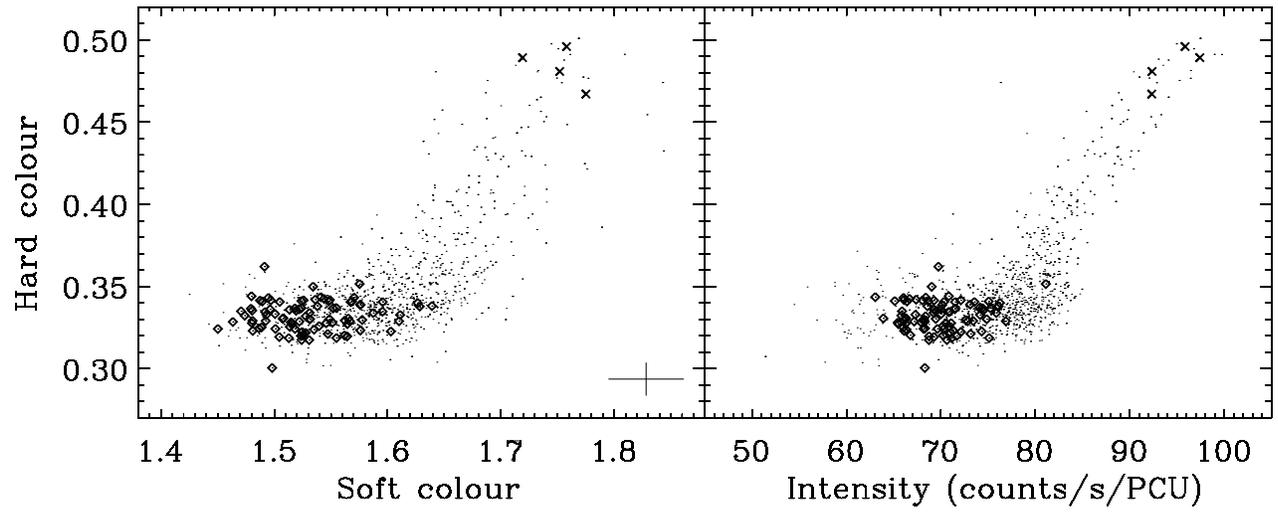}
\caption{Colour-colour diagram (CCD; left panel) and hardness-intensity diagram
(HID; right panel) of the neutron star LMXB XB 1254--690 (same as Fig.~\ref{CCD1}).
The {\it diamond} signs (20.7 ks) and the {\it cross} signs (1.1 ks) 
mark the continuous data sets which were 
used to extract the LB spectrum and the UB spectrum respectively (\S~\ref{SpectralAnalysis}).
\label{CCD4}}
\end{figure*}

We have used the Std2 data of the upper layers of PCU2 in order to do 
the continuum spectral analysis. We have made two spectral files (one for the
LB state, another for the UB state) corresponding
to the largest uninterrupted observational ``good time'' data sets 
after filtering out the 
bursts, dips and flares. Since the LB data set (20.7 ks) is much larger than the
UB data set (1.1 ks; see Fig.~\ref{CCD4}), 
we have first modelled the LB spectrum, which has helped us to decide
on the procedure. Then we have  modelled the UB spectrum following this procedure.
We have done the fitting in the energy range of 2.5--18.0 keV using XSPEC, 
including 1\% systematic instrumental error.

We have used the XSPEC models {\ttfamily bbody}, {\ttfamily diskbb} and 
{\ttfamily comptt} for a blackbody, a multicolour blackbody and a Comptonized 
radiation respectively. In addition, we have used 
{\ttfamily powerlaw}, {\ttfamily cutoffpl} and {\ttfamily bknpower} for
the phenomenological models powerlaw, cut-off-powerlaw and broken-powerlaw 
respectively. 
We have used the last model following the work of \citet{Linetal2007}.
Note that, while the first two models can represent the Comptonized
radiation \citep{RybickiLightman}, the broken-powerlaw model can be an approximation 
for Comptonization under complex conditions or in combination with another
radiation process, such as synchrotron radiation \citep{Linetal2007}.

First, we have checked whether a single component can fit the LB spectrum well.
We have found that {\ttfamily cutoffpl},
multiplied with XSPEC Galactic neutral hydrogen absorption model {\ttfamily wabs},
gives the best fit among all the
single components, and the corresponding $\chi^{2}_{\nu}$ (dof) is $\approx$ 3.23 (33).
This clearly suggests that at least two model components are required to fit the
continuum spectrum. In the two-component scenario, there can be the following 
possibilities: (1) both the accretion disc and the boundary layer are purely
thermal ({\ttfamily diskbb}, {\ttfamily bbody}); (2) the disc emission is thermal,
but the boundary layer emission is Comptonized; (3) the boundary layer emission is 
thermal, but the disc emission is Comptonized; and (4) both of them are Comptonized.
Therefore, for the six XSPEC model components, which we have considered, there can 
be 19 two-component models (see Table~\ref{modellist}). We have fitted the LB 
continuum spectrum with each of these 19 models, multiplied with {\ttfamily wabs} model.
Note that we have frozen the neutral 
hydrogen column density $N_{\rm H}$ of {\ttfamily wabs} to $0.29\times 10^{22}$ cm$^{-2}$ (estimated
with NASA's HEASARC nH tool; see also \citet{Smaleetal2002}) for every model fitting reported in this paper. 
This is because the {\it RXTE} PCA cannot reliably
measure $N_{\rm H}$, which affects the X-ray spectrum mostly at lower energies.
Among these 19 models only six models have given good fits ($\chi^{2}_{\nu} \le 1$),
while for all other models $\chi^{2}_{\nu} \ge 1.5$ (see Table~\ref{modellist}). 
Therefore, we will consider only these six models ({\ttfamily bknpower + diskbb}, 
{\ttfamily comptt + comptt}, {\ttfamily comptt + powerlaw}, 
{\ttfamily comptt + cutoffpl}, {\ttfamily bknpower + comptt}, 
{\ttfamily bknpower + cutoffpl})
for further discussion. Note that, since several two-component models fit
the LB spectrum well, we have not considered more complex models involving
partially thermal and partially Comptonized emission from each of the disc
and the boundary layer. Moreover, addition of an iron line to the spectral model
does not improve the fitting significantly, and does not change the best-fit
parameter values substantially.

\begin{figure*}
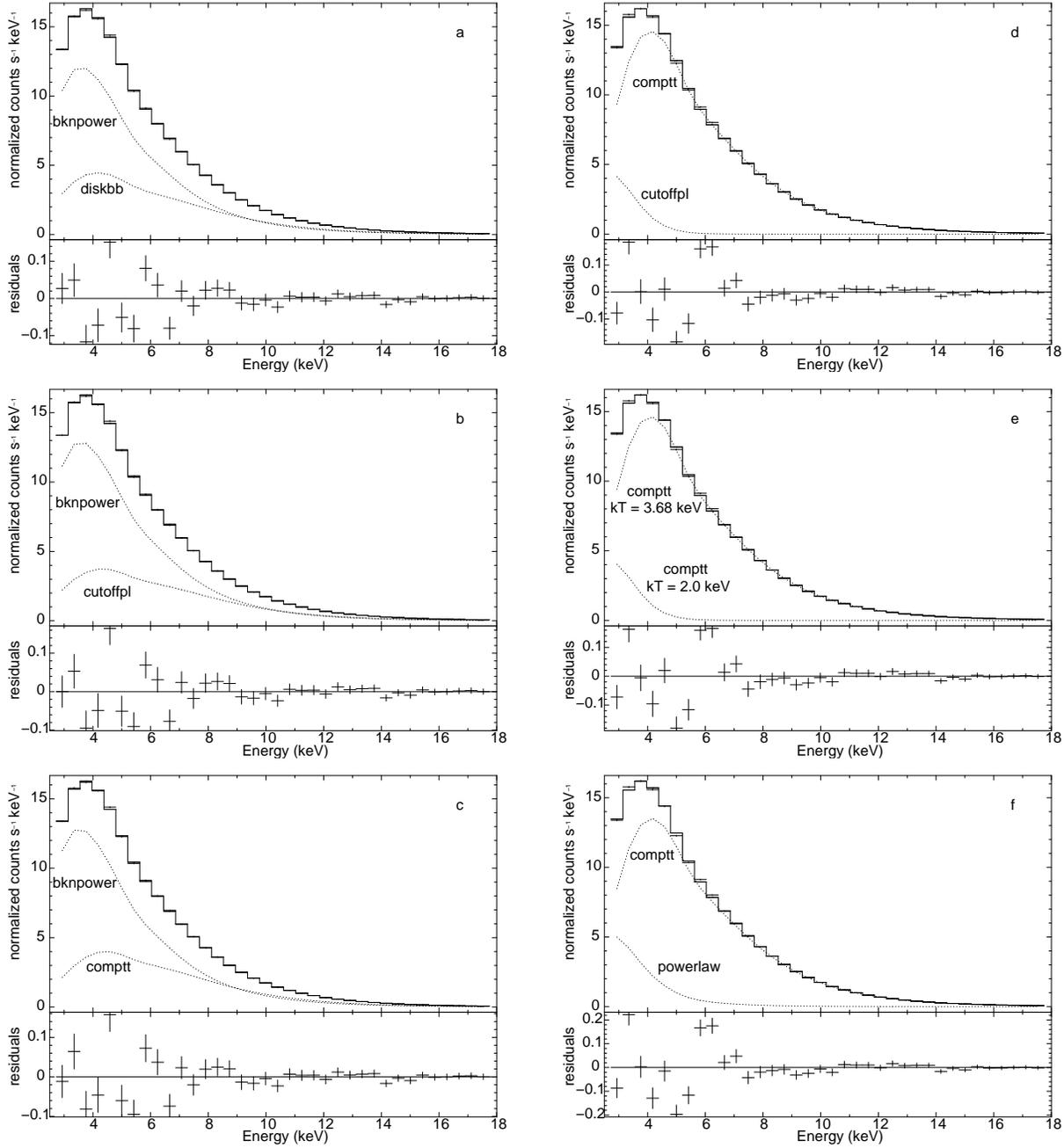

\centering
\begin{tabular}{lr}
\includegraphics*[width=.35\textwidth,angle=270]{diskbbplusbknpower.ps} &
\includegraphics*[width=.35\textwidth,angle=270]{cutoffplpluscomptt.ps} \\
\includegraphics*[width=.35\textwidth,angle=270]{bknpowerpluscutoffpower.ps} &
\includegraphics*[width=.35\textwidth,angle=270]{compttpluscomptt.ps}   \\
\includegraphics*[width=.35\textwidth,angle=270]{bknpowerpluscomptt.ps} &
\includegraphics*[width=.35\textwidth,angle=270]{powerlawpluscomptt.ps} \\
\end{tabular}
\vspace{0.5cm}
\caption {{\it RXTE} PCA continuum spectrum of the LB state of the neutron star LMXB
XB 1254--690. Fitting with six two-component models (see \S~\ref{SpectralAnalysis})
are shown in various panels: (a) {\ttfamily bknpower + diskbb}; (b) 
{\ttfamily bknpower + cutoffpl}; (c) {\ttfamily bknpower + comptt}; (d)
{\ttfamily comptt + cutoffpl}; (e) {\ttfamily comptt + comptt}, (f) 
{\ttfamily comptt + powerlaw}. ``kT" is the plasma temperature of a {\ttfamily comptt}
component. In each panel, the upper sub-panel shows the data points
with error bars, two model component curves and the total model curve, while
the lower sub-panel shows the residuals. The normalizations and shapes of the
model curves suggest that the panels a, b and c (model group-I) 
may represent a particular physical 
picture of the source, while the panels d, e and f (model group-II) may 
represent a different
physical picture (see \S~\ref{SpectralAnalysis} and Table~\ref{LBmodel1}).
\label{specLB}}
\end{figure*}

Table~\ref{LBmodel1} and Fig.~\ref{specLB} suggest that the six best
models can be divided into two groups: the first group consists of 
{\ttfamily bknpower + diskbb}, {\ttfamily bknpower + cutoffpl} and 
{\ttfamily bknpower + comptt}, while the second group consists of 
{\ttfamily comptt + cutoffpl}, {\ttfamily comptt + comptt} and {\ttfamily comptt + powerlaw}.
This is because, 
Table~\ref{LBmodel1} shows that the flux-ratios of the components are close to 2 
for all the models of group-I, with {\ttfamily bknpower} as the dominant model, 
while those are greater than 5.5 for all the models of group-II,  with {\ttfamily comptt}
as the dominant model. Moreover, Fig.~\ref{specLB} clearly shows that
the normalizations and the shapes of the two component-curves of a model are
very similar within a group, but differ significantly between the groups.
Is it now possible to determine which group is preferred?
We have found that the best-fit optical depth ($\approx 10^{-2}$) of
a {\ttfamily comptt} component of the {\ttfamily comptt + comptt} model
is unphysically small (discussed with Lev Titarchuk), and we propose to rule out this model on this ground.
Moreover, the {\ttfamily cutoffpl} and {\ttfamily powerlaw} components of the
{\ttfamily comptt + cutoffpl} and {\ttfamily comptt + powerlaw} models
are likely to represent the ``unphysical" {\ttfamily comptt} component
(see Fig.~\ref{specLB}). Therefore, we argue that each of the three models of group-II 
probably represent the same physical scenario, and hence the entire 
group is not preferred. As a result, we tentatively support the group-I models,
and give the best-fit model parameter values in Table~\ref{LBmodel2}.

Can we now differentiate the three models of group-I? 
The normalization parameter of {\ttfamily diskbb} component is given by
$(R_{\rm in}/D)^2 \cos\theta$, where $R_{\rm in}$ is the apparent inner edge radius
of the disc in km, $D$ is the distance in 10 kpc and $\theta$ is the observer's
inclination angle. 
We have assumed $D = 10$ kpc \citep{DiazTrigoetal2006}, and 
$\theta = 60^{\rm o}-75^{\rm o}$ from the observed dips, and 
lack of eclipses \citep{Motchetal1987, Courvoisieretal1986, Franketal1987}.
The best-fit normalization of the {\ttfamily diskbb} component of 
{\ttfamily bknpower + diskbb} model is 0.27 (Table 3). Therefore, even with
a conservative value of the colour factor ($f = 2.0$), the maximum value
of actual disc inner edge radius comes out to be $\approx 4$ km.
This is less than any realistic value of a neutron star radius, 
and hence is unphysical. We therefore propose to rule out
the {\ttfamily bknpower + diskbb} model. However, this does not necessarily
rule out the {\ttfamily bknpower + cutoffpl} and {\ttfamily bknpower + comptt} models,
because while {\ttfamily diskbb} is a thermal component, {\ttfamily comptt}, as well
as {\ttfamily cutoffpl} (which probably is a phenomenological representation
of {\ttfamily comptt}; see 
Fig.~\ref{specLB}) are Comptonized components. Therefore, we suggest that 
{\ttfamily bknpower + comptt} is the preferred model of LB continuum spectrum of 
XB 1254--690 in the energy range of 2.5--18.0 keV.

\begin{figure}
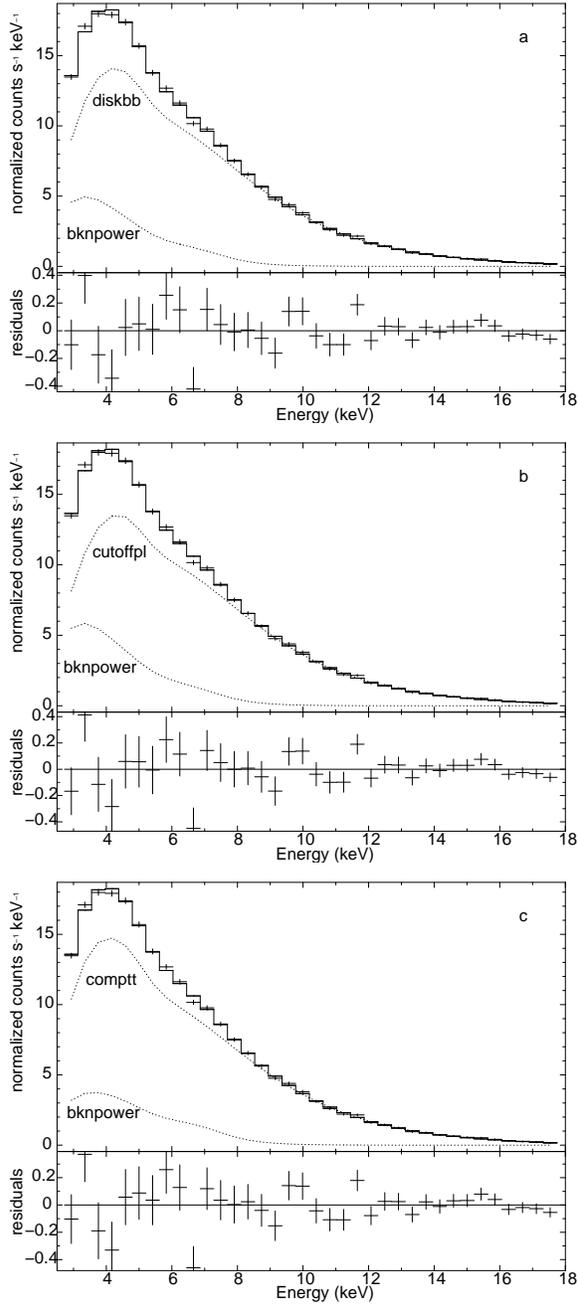

\centering
\begin{tabular}{c}
\includegraphics*[width=.35\textwidth,angle=270]{diskbbplusbknpower_UB.ps}      \\
\includegraphics*[width=.35\textwidth,angle=270]{bknpowerpluscutoffpower_UB.ps} \\
\includegraphics*[width=.35\textwidth,angle=270]{bknpowerpluscomptt_UB.ps}      \\
\end{tabular}
\vspace{0.5cm}
\caption {{\it RXTE} PCA continuum spectrum of the UB state of the neutron star LMXB
XB 1254--690 (similar to Fig.~\ref{specLB}). 
Fitting with three two-component models (see \S~\ref{SpectralAnalysis})
are shown in various panels: (a) {\ttfamily bknpower + diskbb}; (b)
{\ttfamily bknpower + cutoffpl}; (c) {\ttfamily bknpower + comptt}.
The normalizations and shapes of the
model curves suggest that these panels may represent the similar physical
pictures of the source (see \S~\ref{SpectralAnalysis}).
\label{specUB}}
\end{figure}

We have then repeated the above-mentioned procedure for the UB continuum spectrum.
Even for this spectrum, the single component models do not give a good fit, and 
among the 19 two-component models,
the group-I models give the best reduced $\chi^2$: $\chi^2_\nu = 1.31$
for {\ttfamily bknpower + diskbb}, $\chi^2_\nu = 1.36$ for {\ttfamily bknpower + cutoffpl}
and $\chi^2_\nu = 1.39$ for {\ttfamily bknpower + comptt}. In comparison,
the physical model {\ttfamily comptt + comptt} of group-II gives
$\chi^2_\nu = 1.83$. However, in case of UB spectrum, the $\chi^2_\nu$ values for several 
other two-component models are not very different from those for group-I models.
Therefore, it is difficult to identify the correct UB spectral model from the current
data set. Nevertheless, following the results of LB spectrum, and keeping in mind
that the group-I models do give the best $\chi^2_\nu$ values, we suggest that
the group-I models are the preferred models even for UB spectrum. 
Note that, unlike the LB spectrum, 
the {\ttfamily bknpower} component of the group-I models of UB spectrum 
is not the dominant component
(see Fig.~\ref{specUB}). The {\ttfamily diskbb} component of 
{\ttfamily bknpower + diskbb} model suggests an unphysically small disc inner
edge radius ($< 5.4$ km) even for the UB spectrum. Therefore, we propose that
{\ttfamily bknpower + comptt} is the most suitable model to describe the 
UB continuum spectrum of
XB 1254--690 in the energy range of 2.5--18.0 keV. The corresponding best-fit model parameter
values are given in Table~\ref{UBmodel2}. Note that addition of an absorption line or edge 
to the UB spectral model does not affect our conclusions substantially.

%R_in estimate.

\section{Discussion}\label{Discussion}

In this paper, we have studied the timing and spectral properties of the 
neutron star LMXB XB 1254--690. This atoll source was in the banana
state during all the {\it RXTE} observations. We have not detected
burst oscillations from the two bursts in the analyzed data set.
This is not surprising, as only a fraction of bursts from burst oscillation sources
show oscillations \citep{Galloway2008}. However, this somewhat reduces the significance
of the plausible burst oscillation feature reported in \citet{Bhattacharyya2007} 
by increasing the number of trials.
We report the first evidence of QPOs from this source, with frequencies
consistent with the $\sim 30-80$~Hz QPOs typically observed from
atoll sources in intermediate to high states \citep{vanderKlis2006}. 

We have also done the first systematic study of X-ray continuum spectrum
from XB 1254--690. We have fitted an LB spectrum and a UB spectrum with 19
models (see \S~\ref{SpectralAnalysis} and Table 1). In both cases, a 
broken-powerlaw plus a Comptonization model ({\ttfamily bknpower + comptt})
is favoured, and the plasma temperature and the optical depth of the
Comptonization component are consistent with the previously
reported values for other sources \citep{Linetal2007}. 
A broken-powerlaw component was never used to fit the XB 1254--690
continuum spectrum, although it has been recently successfully used for the spectral
fitting of Aql X--1 and 4U 1608--52. This component can represent
Comptonization, either under complex conditions or in combination with 
another radiation process (e.g., synchrotron; see \S~\ref{SpectralAnalysis}).
These imply that the blackbody radiations from the disc and the boundary
layer are entirely processed before reaching the observer. Note that
the classical models {\it Eastern} and {\it Western} 
give a much worse fit than the {\ttfamily bknpower + comptt} model.
For example, $\chi_\nu^2$ (dof) values for 
{\it Eastern}, {\it Western} and {\ttfamily bknpower + comptt} models are
1.96 (30), 1.50 (30) and 0.54 (28) respectively for the LB spectrum.
Besides, the {\ttfamily cutoffpl + bbody}
and {\ttfamily diskbb + powerlaw} models, which were previously used for XB 1254--690
\citep{Smaleetal2002, Iariaetal2007}, give much worse 
fits ($\chi_\nu^2$ (dof): 1.65 (31) and 1.95 (32) respectively for LB).
However, although a broken-powerlaw component gives the best fit, note that
it is only a phenomenological model, and should be replaced with a 
physical model in the future in order to clearly understand the meaning of 
the best-fit parameter values. A physical model will also be necessary
to interpret the high $\Gamma_2$ of the UB broken-powerlaw (Table \ref{UBmodel2}),
which indicates a quick cut-off after the first powerlaw component.

With the best-fit model in hand, we have attempted to determine the relative
energy budgets of the disc and the boundary layer. For the LB spectrum,
the flux from the {\ttfamily bknpower} component is about 2.1 times
larger than the {\ttfamily comptt} component (Table 2). Therefore, assuming
one component entirely originates from the disc and the other component
fully comes out of the boundary layer, we can make the following 
conclusions using the Fig. 5 of \citet{Bhattacharyyaetal2000}.
(1) The spin frequency depends on the chosen
EoS model of the neutron star, with larger frequencies
for softer EoS models. Therefore, an independent measurement of the
stellar spin frequency can be useful to constrain the EoS models.
(2) If the {\ttfamily bknpower} component originates from the disc, then
the neutron star is spinning with a frequency close to the break-up 
frequency (irrespective of the EoS model).
However, these simple conclusions may not be reliable for one or more
of the following reasons.
(1) Each of the spectral components may originate partially from
the disc, and partially from the boundary layer.
(2) Disc emission and/or boundary layer emission may be partially
obscured.
(3) Boundary layer emission (and also the inner disc emission) 
may be partially reprocessed by the disc.
Besides, the {\ttfamily bknpower} flux is 
smaller than the {\ttfamily comptt} flux
for the lower quality UB spectrum, although the spectral
shapes and parameters of these two components are very similar
to those of the LB spectrum (see Tables 3 and 4; Figs.~\ref{specLB} and
\ref{specUB}). This difference in the normalization ratio of the 
two components may be due to one or more of the above three reasons.
One main problem in identifying and understanding the emissions from
various source components is the lack of simultaneous broadband (optical, 
UV and X-ray) data. The optical and UV emissions are expected to originate
by the reprocessing of the X-ray photons at the outer disc, and 
hence the correlations among time variations of spectral properties
(e.g., \citet{Balucinskaetal2004}), especially
in different wavelengths will significantly contribute to the understanding 
of the neutron star LMXB components. The proposed {\it Astrosat} 
satellite will be useful for such a study in the near future.

\section*{Acknowledgments}

We thank Tod E. Strohmayer for encouragement, Lev Titarchuk for a useful discussion, and an anonymous referee for
constructive comments which improved the paper.
This work was supported in part by the US NSF grant AST 0708424.

\clearpage
% use packages: array
\begin{table}
 \centering
\caption{A list of two-component models used to fit the {\it RXTE} PCA continuum spectra of
the neutron star LMXB XB 1254--690 (\S~\ref{SpectralAnalysis}).}
\begin{tabular}{cccc}
\hline
Sr. No. & Fitting models\footnotemark[0] & $\chi^{2}_{\nu}$ (dof)\footnotemark[1] & Number of \\
& & & free parameters \\
\hline
1     & diskbb\footnotemark[2] + comptt	& 1.95 (30)	    & 6				\\
2     & diskbb + powerlaw	   		& 1.95 (32)	    & 4				\\
3     & diskbb + cutoffpl\footnotemark[3]	& 1.97 (31)	    & 5				\\
4     & diskbb + bknpower\footnotemark[4]	& 0.52 (30)	    & 6				\\
5     & comptt\footnotemark[5] + bbody		& 1.50 (30)	    & 6				\\
6     & powerlaw + bbody\footnotemark[6]	& 2.24 (32)	    & 4				\\
7     & cutoffpl + bbody	   		& 1.65 (31)	    & 5				\\
8     & bknpower + bbody	   		& 2.39 (30)	    & 6				\\
9     & diskbb + bbody		   		& 2.75 (32)	    & 4				\\
10    & comptt + comptt	   		& 0.91 (28)	    & 8				\\
11    & powerlaw + comptt	   		& 0.95 (30)	    & 6				\\
12    & cutoffpl + comptt			& 0.91 (29)	    & 7				\\
13    & bknpower + comptt	  		& 0.54 (28)	    & 8				\\
14    & powerlaw + powerlaw	   		& 32.31 (32)	    & 4				\\
15    & cutoffpl + powerlaw			& 2.33 (31)	    & 5				\\
16    & bknpower + powerlaw			& 3.47 (30)	    & 6				\\
17    & cutoffpl + cutoffpl		 	& 1.54 (30)	    & 6				\\
18    & bknpower + cutoffpl		 	& 0.53 (29)	    & 7				\\
19    & bknpower + bknpower		 	& 3.73 (28)	    & 8				\\
\hline
\end{tabular}
\begin{flushleft}
$^0$All models have been multiplied with the {\ttfamily wabs} absorption model of XSPEC,
with the fixed  $N_{\rm H} = 0.29 \times 10^{22}$cm$^{-2}$. \\
$^1$Reduced $\chi^2$ (degrees of freedom) for the LB spectrum \\
$^2$Multicolour blackbody model of XSPEC. \\
$^3$A powerlaw model with high energy exponential cut off. \\
$^4$A broken power law with break energy $E_{\rm break}$. \\
$^5$A Comptonization model in XSPEC. \\
$^6$Blackbody model in XSPEC. \\
\end{flushleft}
\label{modellist}
\end{table}

\clearpage
%\end{center}
%\baselineskip=22pt
%\pagestyle{plain}
%\thispagestyle{empty}
%\null
%\normalsize
\begin{table*}
 \centering
\caption{Flux values of XB 1254--690 from fitting of the LB continuum spectrum with 
the six two-component models, which give $\chi^2_\nu \le 1$.}.
\begin{tabular}{|c|ccc|ccc|}
\hline 
\multicolumn{1}{|c}{} &
	\multicolumn{3}{|c|} {Model Group-I} &
	\multicolumn{3}{|c|} {Model Group-II} \\
\cline{2-7}
Models                              & bknpower & bknpower  & bknpower & comptt & comptt   & comptt   \\                              & +diskbb  & +cutoffpl & +comptt  & +cutoffpl  & +comptt  & +powerlaw    \\
\hline
{$\chi^{2}_{\nu}$ (dof)}                  & 0.52(30) & 0.53(29)  & 0.54(28) & 0.91(29) & 0.91(28) & 0.95(30)   \\
{Absorbed total flux}\footnotemark[1]     & 7.88     & 7.88      & 7.90     & 7.96     & 8.00     & 7.96       \\
{Unabsorbed total flux}\footnotemark[1]   & 8.02     & 8.02      & 8.04     & 8.11     & 8.15     & 8.11       \\
{Flux of the first component}\footnotemark[1]           & 5.28     & 5.60      & 5.44     & 7.25     & 7.27   & 6.88     \\
{Flux of the second component}\footnotemark[1]           & 2.74     & 2.42      & 2.60     & 0.86     & 0.87    & 1.23   \\
{Ratio of the two fluxes} & 1.92     & 2.31      & 2.09     & 8.42     & 8.45     & 5.59       \\
\hline
\end{tabular}
\begin{flushleft}
$^1$In the range 2.5--18.0 keV and in unit of $10^{-10}$ ergs cm$^{-2}$ s$^{-1}$.
\end{flushleft}
\label{LBmodel1}
\end{table*}

\clearpage
% use packages: array
\begin{table*}
 \centering
\caption{Spectral fitting parameters of the LB state of XB 1254--690 for 
the three two-component models (see Table~\ref{LBmodel1} and \S~\ref{SpectralAnalysis}).
Best-fit values are accompanied with 90\% errors.}
\begin{tabular}{cccccccc}
\hline
Sr. No. & Model             & Component   & Parmeters 		& Best-fit values 	  	& Component   & Parmeters 		& Best-fit values 	  \\ 
\hline
1    & bknpower\footnotemark[1]+diskbb   & bknpower & $\Gamma_{1}$\footnotemark[1]  		&$2.40_{-0.091}^{+0.17}$	& diskbb   & Tin\footnotemark[2]&$2.82_{-0.34}^{+0.20}$   \\ 
       &                   		   &          & $E_{\rm break}$\footnotemark[1]	&$6.29_{-0.18}^{+0.18}$         &          & norm      		&$0.27_{-0.062}^{+0.16}$  \\ 
       &                   		   &          & $\Gamma_{2}$\footnotemark[1]  		&$3.85_{-0.39}^{+0.83}$         &          &           		&  	                  \\ 
       &                   		   &          & norm      		&$0.43_{-0.042}^{+0.057}$       &          &           		&      	                  \\ 
\hline
2   & bknpower+cutoffpl\footnotemark[3] & bknpower & $\Gamma_{1}$  		&$2.43_{-0.10}^{+0.33}$   	& cutoffpl & $\alpha$\footnotemark[3]  		&$-0.36_{-0.36}^{+1.1}$   \\ 
       &                   		   &          & $E_{\rm break}$  &$6.25_{-0.22}^{+0.31}$ 	&          & $\beta$\footnotemark[3]&$3.00_{-1.2}^{+1.4}$\\ 
       &                   		   &          & $\Gamma_{2}$  		&$3.83_{-0.36}^{+1.6}$		&          & norm      		&$0.011_{-0.011}^{+0.025}$\\
       &                   		   &          & norm     		&$0.48_{-0.14}^{+0.066}$	&          &           		&			  \\ 
\hline
3  & bknpower+comptt   & bknpower & $\Gamma_{1}$  		&$2.52_{-2.2}^{+0.80}$		& comptt   & T0\footnotemark[4]	&$1.12_{-1.1}^{+1.1}$	  \\ 
       &                   &          & $E_{\rm break}$  &$6.26_{-0.26}^{+0.56}$		&          & kT\footnotemark[5] &$3.00_{-2.6}^{+7.7}$	  \\ 
       &                   &          & $\Gamma_{2}$  		&$3.89_{-0.45}^{+5.3}$		&          & taup\footnotemark[6]&$6.82_{-6.8}^{+183}$	  \\ 
       &                   &          & norm      		&$0.48_{-0.38}^{+1.2}$		&          & norm      		&$0.02_{-0.013}^{+0.023}$ \\ 
\hline
\end{tabular}
\begin{flushleft}
$^1$A broken power law: \\
$$
A(E) = KE^{-\Gamma_{1}} \;\;  \mathtt{ for } \;\;  E \le E_{\rm break};
$$
$$
= KE_{\rm break}^{\Gamma_2 - \Gamma_1} \times (E/1keV)^{-\Gamma_{2}}\;\;  \mathtt{ for } \;\; E \ge E_{\rm break}.
$$
$E_{\rm break}$ is in keV. \\
$^2$Temperature at inner disc radius (in keV). \\
$^3$A power law with high energy exponential cut off:
$$A(E) = KE^{-\alpha} exp{(-E/\beta)}.
$$
$\beta$ is in keV. \\
%K(norm)=photons cm$^{-2}$ keV$^{-1}$ s$^{-1}$ \\
%HighECut($\beta$) = e-folding energy of exponential rolloff (in keV) \\
$^4$Input soft photon (Wien) temperature (in keV). \\
$^5$Plasma temperature (in keV). \\
$^6$Optical depth of the corona. \\
\end{flushleft}
\label{LBmodel2}
\end{table*}

\clearpage
% use packages: array
\begin{table*}
\centering
\caption{Spectral fitting parameters of the UB state of XB 1254--690 for
a two-component model (see \S~\ref{SpectralAnalysis}). Best-fit values are accompanied with 90\% errors.}
\begin{tabular}{ccccccc}
\hline
Model             & Component    & Parmeters\footnotemark[1] 		& Best-fit values	& Component    & Parmeters\footnotemark[1] 	    & Best-fit values\\ 
\hline 
bknpower+comptt   & bknpower & $\Gamma_{1}$  		&$2.32_{-0.32}^{+0.35}$	& comptt   & T0 &$0.67_{-0.06}^{+0.07}$\\ 
                 &          & $E_{\rm break}$ &$7.22_{-0.97}^{+0.40}$	&          & kT  &$2.79_{-0.12}^{+0.18}$\\ 
                  &          & $\Gamma_{2}$\footnotemark[2]  		&$9.99_{-0.1}^{+0.008}$	&          & taup &$7.78_{-1.7}^{+2.2}$	\\ 
                  &          & norm      		&$0.12_{-0.07}^{+0.22}$	&          & norm		 &$1.08_{-0.4}^{+0.1}$	\\ 
\hline
\end{tabular}
\begin{flushleft}
$^1$Definitions and units of parameters are same as in Table~\ref{LBmodel2}.\\
$^2$The large value of the photon index indicates a quick cut-off after the first
powerlaw component (see \S~\ref{Discussion}).\\
\end{flushleft}
\label{UBmodel2}
\end{table*}

\clearpage

%\clearpage
%\begin{figure*}
%\centering
%\includegraphics*[width=\textwidth]{CCDandHIDplot_BURST-DIP-FLARE_48by20-final.ps}
%\caption{Two bursts (symbol-crosses), dips (symbol-diamonds) and flares (symbol-triangles) are marked in CCD (left panel) and in HID (right panel).
%\label{CCD2}}
%\end{figure*}

\clearpage

\clearpage

\clearpage

\topmargin -0.125in
\oddsidemargin -0.6in
\evensidemargin 0in
\textheight 24cm
\textwidth 16cm
\parskip 1ex    % White space between paragraphs amount
\raggedbottom

\clearpage

\clearpage


\begin{thebibliography}{99}

\bibitem[Balucinska et al. (2004)]{Balucinskaetal2004} Baluci\'nska-Church M., Church M.J., 
Szostek A., 2004, NuPhS, 132, 576

\bibitem[Barret (2001)]{Barret2001} Barret D., 2001, Advances in Space
Research, 28, 307

\bibitem[Bhattacharyya (2002)]{Bhattacharyya2002} Bhattacharyya S., 2002,
A\&A, 383, 524

\bibitem[Bhattacharyya (2007)]{Bhattacharyya2007} Bhattacharyya S., 2007, MNRAS, 377, 198

\bibitem[Bhattacharyya (2010)]{Bhattacharyya2010} Bhattacharyya S., 2010,
Advances in Space Research, 45, 949

\bibitem[Bhattacharyya et al. (2000)]{Bhattacharyyaetal2000} Bhattacharyya S., Thampan
A.V., Misra R., Datta, B., 2000, ApJ, 542, 473

\bibitem[Chakrabarty et al. (2003)]{Chakrabartyetal2003} Chakrabarty D., Morgan
E.H., Muno M.P., Galloway D.K., Wijnands R., van der Klis M., Markwardt C.B.,
2003, Nature, 424, 42

\bibitem[Christian \& Swank (1997)]{ChristianSwank1997} Christian D.J., Swank J.H., 
1997, ApJS, 109, 177

\bibitem[Church \& Baluci\'nska-Church (2001)]{ChurchBalucinskaChurch2001}
Church M.J., Baluci\'nska-Church M., 2001, A\&A, 369, 915

\bibitem[Courvoisier et al. (1986)]{Courvoisieretal1986} Courvoisier T.J.-L., Parmar A.N., Peacock A., Pakull M., 1986, ApJ, 309, 265

\bibitem[D\'iaz Trigo et al. (2006)]{DiazTrigoetal2006} D\'iaz Trigo M., 
Parmar A.N., Boirin L., M\'endez M., Kaastra J.S., 2006, A\&A, 445, 179

\bibitem[Frank et al. (1987)]{Franketal1987} Frank J., King A.R., Lasota J.-P., 
1987, A\&A, 178, 137

\bibitem[Galloway et al. (2008)]{Galloway2008} Galloway D.K., Muno
M.P., Hartman J.M., Psaltis D., Chakrabarty D.,
2008, ApJSS, 179, 360

\bibitem[Gilfanov et al. (2003)]{Gilfanovetal2003} Gilfanov M., Revnivtsev M., 
Molkov S., 2003, A\&A, 410, 217

\bibitem[Griffiths et al. (1978)]{Griffithsetal1978} Griffiths R.E., Gursky H., Schwartz D.A., Schwarz J., Bradt H., Doxsey R.E., Charles P.A., Thorstensen J.R., 1978, Nature, 276, 247

\bibitem[Iaria et al. (2001)]{Iariaetal2001} Iaria R., Di Salvo T., Burderi L.,
Robba N.R., 2001, ApJ, 548, 883

\bibitem[Iaria et al. (2007)]{Iariaetal2007} Iaria R., Di Salvo T., 
Lavagetto G., D'A\'i, A., Robba N.R., 2007, A\&A, 464, 291

\bibitem[Lin et al. (2007)]{Linetal2007} Lin D., Remillard R.A., Homan, J.,
2007, ApJ, 667, 1073

\bibitem[Maccarone \& Coppi (2003)]{MaccaroneCoppi2003} Maccarone T.J., Coppi P.S.,
2003, A\&A, 399, 1151

\bibitem[Maitra \& Bailyn (2004)]{MaitraBailyn2004} Maitra D., Bailyn C.D., 2004,
ApJ, 608, 444

\bibitem[Mason et al. (1980)]{Masonetal1980} Mason K.O., 
Middleditch J., Nelson J.E., White N.E., 1980, Nature, 287, 516

\bibitem[Mitsuda et al. (1989)]{Mitsudaetal1989} Mitsuda K., Inoue H., 
Nakamura N., Tanaka Y., 1989, PASJ, 41, 97

\bibitem[Motch et al. (1987)]{Motchetal1987} Motch C., Pedersen H., Beuermann H.,
Pakull M.W., Courvoisier T.J.-L., 1987, ApJ, 313, 792

\bibitem[Olive et al. (2003)]{Oliveetal2003} Olive J.-F., Barret D., Gierli\'nski, M.,
2003, ApJ, 583, 416

\bibitem[Rybicki \& Lightman (1979)]{RybickiLightman} Rybicki G.B., Lightman A.P. 1979,
{\it Radiative Processes in Astrophysics}, John Wiley \& Sons, New York

\bibitem[Smale et al. (2002)]{Smaleetal2002} Smale A.P., Church M.J., 
Baluci\'nska-Church M., 2002, ApJ, 581, 1286

\bibitem[Strohmayer and Bildsten (2006)]{StrohmayerBildsten2006}
Strohmayer T.E., Bildsten L., 2006, in {\it Compact
Stellar X-ray Sources}, eds. Lewin W.H.G., van der Klis M.,
Cambridge Univ. Press, 39, 113

\bibitem[van der Klis (2006)]{vanderKlis2006} van der Klis M., 2006, 
in {\it Compact Stellar X-ray Sources}, eds. Lewin W.H.G., van der Klis M., 
Cambridge Univ. Press, 39, 39 

\bibitem[White et al. (1988)]{Whiteetal1988} White N.E., Stella L., Parmar
A.N., 1988, ApJ, 324, 363

\bibitem[Wijnands (2001)]{Wijnands2001} Wijnands R., 2001, Advances in Space
Research, 28, 469

\end{thebibliography}
\end{document}